# Comparative Study of Cloud and Non-Cloud Gaming Platform: Aperçu

Prerna Mishra* and Urmila Shrawankar**
Assistant Professor*
J. D. College of Engineering, Nagpur (MS), India
prerna.mishra.891@gmail.com

Associate Professor **
Department of Computer Science & Engineering,
G. H. Raisoni College of Engineering, Nagpur (MS), India
urmila@ieee.org

**Abstract:** Nowadays game engines are imperative for building 3D applications and games. This is for the reason that the engines appreciably reduce resources for employing obligatory but intricate utilities. This paper elucidates about a game engine and its foremost elements. It portrays a number of special kinds of contemporary game engines by way of their aspects, procedure and deliberates their stipulations with comparison.

**Keywords**: Game Engine, Non-Cloud Gaming Platforms, Cloud Gaming Platforms, Platform Popularity, 2D and 3D Engines.

## 1. Introduction:

In general, the notion of game engine [1] is very easy to comprehend. It is a platform for performing game related tasks like interpretation, physics related reckoning, and to facilitate developers for focusing on the niceties that make the game inimitable. Engines are in reality an assemblage of reusable modules that can be manipulated in order to carry a game towards realism. Indeed, there are certain disparities between a game and a game engine [2]. Graphics, animation, audio, physics, UI and AI are the major different constituents of an engine. Conversely the subject matter of a game, its definite characters and background, real world avatar and its behaviours etc. are the components that create the real game. Game engines are middleware's.

Game engines produce the replication of actual world in the digital world by controlling the elementary physics. Games developed by these engines make user, casual or die-hard player. In general, smartphone operators, as mobile game players, are alike casual gamers rather than die-hard gamers. In contradiction of die-hard gamers, casual gamers are outlined as less dedicated, less spirited, and more tranquil users. Casual gamers incline to be not as much of ardent and less fascinated to classy or multifarious games (comparative to gameplay, achieve target, environs, graphics, chaps, etc.).The market for video games is growing, with sales in 2015 of $91.5 billion marking an 11.84% increase over 2014, at this price global revenues are expected to reach $107 billion in 2017. However, growth is not only in sales but also in the miscellany of matter offered, vacillating from scholastic games to first-person shooters. In addition, a captivating conjunction of mass media is proceeding with video games, having motion picture eminence cut-scenes and voiceover.

In a cloud environment, the task of service provider is separated into two: the infrastructure providers managing cloud platforms and rent assets conferring to a usage, and service providers leasing assets from one or more infrastructure providers to assist the users. Lately, a novel type of cloud service has been familiarized, which have the utmost severe exigencies on network Quality of Service (QoS) to date known as cloud gaming. However, in cloud gaming the complete user experience is provided through the network. This creates dissimilarity between cloud gaming and conventional online Gaming in stipulations of network quality of experience (QoE). While in conformist Online Gaming the user experience is spawned at the client side so the network does not have any impact on the performance, affecting the worth of Cloud Gaming[3]. With the constituent of video gaming disappearing, people merely favour to finance time in real-time mobile games that are compatible with an extensive array of platforms and operating systems. These engines takes the gaming experience to an entirely new-fangled echelon, avoiding poor graphics and quality experiences with the similar joysticks from the past to play around.

This paper portrays the comparative study offers diverse cloud and non-cloud platforms that are currently associated with gaming. As games are evolving progressively delivering profounder and added abiding experiences for players, their prospective for psychosomatic impression is growing in proportion. Some renowned companies of games are Microsoft Game Studios, Electronic Arts, etc [4].Console producers [4] are a company yielding and disseminating video game consoles. Some of the utmost familiar console producers are Atari, Microsoft Corporation, Nintendo Company, Sega, and Sony Computer Entertainment Inc. However, currently there are three major popular platforms Microsoft Xbox 360, Sony PlayStation3 and Nintendo apart from mobile gaming apps as shown in figure 1.

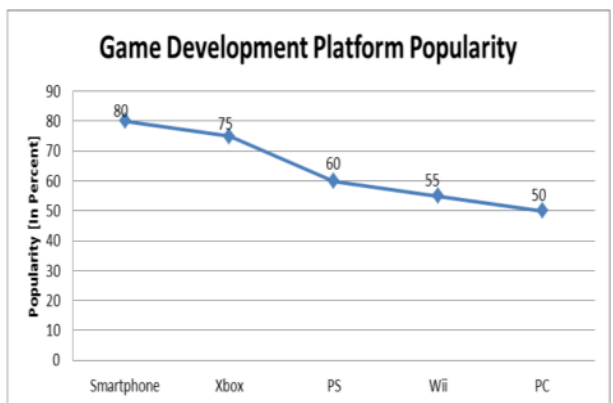

Figure 1. Popularity of gaming platforms in India

## 2. Comparative study:

### 2.1. Non-Cloud Platform:

Analysis amid numerous game engines is a daunting errand for the reason that of their innumerable field, kinds, hypermedia sustenance, middleware support, language and platform enslavements and many other deputized characteristics. It has to be acknowledged that the non-cloud game platforms have dissimilar features, modules, benefits and restrictions relating some eminent game engines. Six popular game engines are compared [5] with each other namely Cry Engine 3, Hero Engine, Source 2 Engine, Unity 4 Engine, Unreal Engine 4 and Vision Engine 8.Table 1 expresses the efficacy in terms of platform reliance, interface and language, intrinsic physics and AI engine supported with forward and backward compatibilities.

An analysis was completed with the help of 15 students (beginner coders) and 20developers from different game industries who gave their views on a number of game engines currently accessible in the marketplace in India. Figure 1 depicts the popular gaming platforms in India.

### 2.2. Cloud Gaming Platforms

A cloud gaming structure assembles the participant's activities, transfers it to the cloud server, concocts the act, extract the consequences, conceals the resultant modifications into the gaming simulating environment and rivulets the gaming sequences in return to the player. For safeguarding inter communication, all these quential actions tends to happen in a period of milliseconds. Subliminally, the total time, which can be said as communication interruption or delay, essentially are kept negligibly conceivable for delivering an amusing involvement to cloud game players. One of the modest methodologies for supporting cloud gaming is to use the services of general desktop streaming thin clients, such as Ubitus [12], GamingAnywhere [13], and VirtualGL [14]. Cloud gaming not only necessitates elevated continual downlink bandwidth but low latency too [6]. Cloud gaming by now has engendered a prodigious contract of curiosity amongst business persons, venture investors, and the end users. Some prominent cloud platforms are discussed below.

*A.  Gaikai*

In GaiKai [7], games are uploaded to the datacenters situated largely around the globe. From these centers, they are cascaded using high-end servers to the devices connected to each other by medium of internet, analogous to the manner videos are streamed to the user's computer. It endows the capability of streaming the graphically rich and real games and additional related data rapidly to more or less on any devices from anywhere in the world. Gaikai has developed the utmost excellence; quickest communicating cloud-streaming platform on the globe, aiding a platform to developer's proficient enough for providing games and other communicating matter instantaneously to the end users via the Internet. Nevertheless Gaikai do not sustain packages on devices like digital TVs and tablets [8].

*B.  StreamMyGame*

StreamMyGame [7][10] is a wide-ranging software elucidation that empowers games and applications to be played remotely. One accesses and plays games tenuously via their local/home network. This gaming platform gains access and engages in gameplay tenuously by means of their broadband network, note downs gameplay to High definition Video archives, one can upload recorded HD video files to any online video sites and publicize games, so anybody on their native system can perceive it and participants gets access to amenities of meeting other players by means of creating the groups with forums, chats, recognize new and former participants presence.

*C.  OnLive*

OnLive [7] [9] conveys on request real-time communal experiences with opulent content via the Internet. OnLive uses cloud to deliver the potency and astuteness needed to immediately convey comprehensive applications having animations, graphics, AI, physics, etc. OnLive make use of virtual machines on customized servers with graphics processing units (GPUs) and commercial compression procedures, which has to work-out two issues for game respectively. Live stream is augmented for gameplay depending on physical-world Internet circumstances and Media stream is a server-side full HD stream managing viewers or players for recording and reviewing progressions of their games. In spite of the benefits, OnLive has quite a few restrictions. Actually, OnLive is no treasonably noble enough for accomplishing1080p resolution. That is to say for the reason that it requires broadband speediness of up to 10 MB per second.

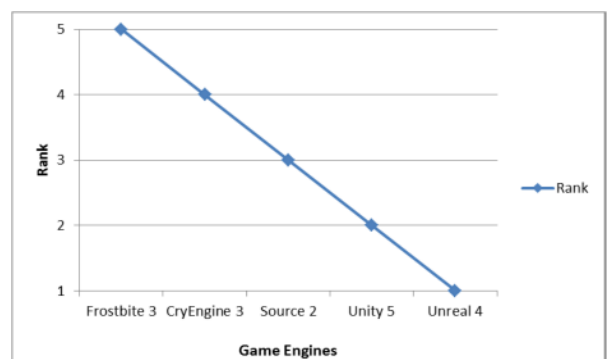

Figure 2. Best Game Engines for Developers as per usability

*D.  GamingAnyWhere*

GamingAnyWhere is the paramount open-source game development platform [7] [11]. It concedes scholars to experiment notions and concepts on a cloud gaming proving-ground, gaming bringers to cultivate services on it and players can fabricate their personal gaming clouds from their computers [11]. Moreover, of its candidness, it is designed for high affability, maneuverability, and reassessability. Platforms supported by GamingAnywhere are Windows, Linux, and OS X, iOS and Android. GamingAnyWhere is multi-platform, proficient in footings of time and space complexities and offers accessible unrestricted platform.

Table 1. Assessment of engines on performance aspects

| Game Engine | Platforms | Language Support | AI Engine | Physics Engine | Forward compatibility | Backward compatibility |
|---|---|---|---|---|---|---|
| CryEngine 3 | Win, X360, PS3, Wii U | C++, Visual Script, Lua | Lua-driven AI | Soft-body | No | Yes |
| Hero Engine | Win | Hero Script | AIseek | PhysX | Partial | Yes |
| Source 2 Engine | Win, Mac, Xbox 360, Wii, Linux, android | C++ | AI Director | Ipion | No | Partial |
| Unity 4 | BlackBerry, Win Phone, Win, OS X, Android, iOS, Apple TV, PS3/4, PS Vita, Xbox 360, Xbox One, Wii U, Wii. | C#, JavaScript, Boo | RAIN | PhysX | Partial | Yes |
| Unreal 4 Engine | Windows, OS X Linux, Xbox 360/One, PS3/4, Wii U, Android, iOS, WinRT, PS Vita | C++, C#, GLSL, CG, HLSL | Kynapse | PhysX | Partial | Yes |
| Vision Engine 8 | Windows, Xbox 360, PS3, Wii, Wii U, iOS, Android, Win Phone, PS Vita | C++ | Kynapse | Bullet, ODE, PhysX | No | Partial |

### 2.3. Unity Vs. Unreal

The two most popular game engines: Unity and Unreal are the best engines for developing 2D and 3D games but have certain limitations too. Unity has a very persuasive civic of resource and plug-in makers. It provisions a widespread array of platforms, mobile, desktop, web and gaming consoles. Its 3D engine produces results of high eminence deprived of any multifaceted arrangement. There are a small number of concerns which are worth cogitating before selecting Unity. Collaboration in Unity is challenging, Performance is not enough pronounced till very recent Unity version is processed entirely on a solitary thread and avoids the usage of the additional cores in almost every mobile devices. Unity 5 is surmounting these challenges.

The Unreal Engine has a long history as one of the top 3D games engines for PC and console platforms. The Unreal Engine is AAA game quality having incredible performance producing the best and most realistic graphics. State of the art tools are amalgamated in to entirely stages of game development. It is open i.e. provides wholesome access of source code letting expansion, customization and bug fixation. The Unreal Engine is an outstanding preference for high class 3D games on high end mobile devices but it is not for everyone. It is scripted in C++, thus edgy for beginner developers.

### 3. From Advancements to The Future of Gaming Technology

Video games has emanated from a long way ever since they overlapped into the conventional in the 1980s, but some incredible improvements in the technologies made the future of gaming even dazzling. 3D scanning and facial recognition technology has permitted developers to actually develop their likeness in the game or to ingeniously transfer their expressions to additional digital creations. For example, Intel's RealSense 3D camera allows gaming developers to produce games adapting the emotions of the gamer by scanning different locations of a player's face. Voice controlled gaming has remained all over the place for a while, but the prospective of using this technology in gaming systems has finally captivated to realism, now even computers are capable to effortlessly diagnose voice commands from the player. Player can now turn the game console on and off, supervise gameplay, interact on social game group, all by commanding to your gaming system.

Using a 3D camera that trails distinct area of your body, gesture control consents players to connect to their gaming experience by means of the usual body movements. State of the art advancements deliver to players a rich and high quality experience in form of fully rendered worlds with photo realistic textures.

Whether it's smart watches or goggles, wearable games make gaming mobile without being too intrusive. Wearable games are extensions of the body in addition to extensions of the consoles used while playing. With the dawn of smartphone's, the entire gaming experience is now in the hand of the players. Instead of developing gaming structures necessitating more influential hardware, developers have lessened the load with the use of cloud computing. By usage of clouds, game streaming is now a reality, streaming similar to movie streaming.

However many Virtual Reality (VR) gaming consoles are not released commercially yet, but this developing VR technology is dignified to concede gamers a wholly immersive gaming experience. VR are sometimes stated as immersive multimedia or computer-simulated reality duplicates environs by simulating a corporeal existence at places in the physical-world or a fantasy world, letting the players to communicate with that world within the simulated environment. Virtual Reality is imminent. It's no overstatement to state that VR is one of the gigantic advancements in gaming world and eradicating the blockade in the midst of player and game world.

## 4.   Conclusion:

From the commencement of game engines to the latest 3D jazzed-up game engines, the objective of development was to endure the equivalent i.e. giving game coders a rostrum for creating their unique games into reality. For them it is not necessary to code or build the game from the scuff however hypothetically execute the idea aided by some popular game engines. These engines offer the rudimentary central design with codes and the manifestation as a middleware. A game developer solely requires tugging them as per their own peculiar necessity. The progression of gaming engines is at the present are proceeding en route for supplementary realistic and technically sound games in innumerable grounds like physics, sounds, AI, graphics, and animations etc.

### Acknowledgment:
We wish to thank all the students and game developers around the country who actively contributed in this comparative analysis.